\documentclass[11pt]{article}
\usepackage[dvips]{graphicx}
\usepackage{psfrag}
\usepackage{amssymb,amsthm}
\usepackage{amsmath}
\usepackage{amscd}
\usepackage{latexsym}

\def\beq{\begin{equation}}
\def\eeq{\end{equation}}
\def\bea{\begin{eqnarray}}
\def\eea{\end{eqnarray}}

\catcode`@=11
\renewcommand{\section}{\@startsection{section}{1}{0pt}{\medskipamount}
{\medskipamount}{\large\bf}}
\numberwithin{equation}{section}
\catcode`@=12
\def\a{\alpha}
\def\b{\beta}
\def\g{\gamma}

\def\de{\delta}
\def\ve{\varepsilon}
\def\h{\eta}
\def\t{\theta}

\def\la{\lambda}
\def\m{\mu}
\def\L{\Lambda}
\def\n{\nu}

\def\s{\sigma}

\def\vp{\varphi}

\def\c{\chi}
\def\ome{\omega}
\def\Om{\Omega}
\def\Ga{\Gamma}

\def\gt{\tilde g}
\def\et{\tilde e}
\def\ot{\tilde\omega}

\def\oh{\widehat\omega}
\def\1{{\bar 1}}
\def\2{{\bar 2}}
\def\3{{\bar 3}}
\def\4{{\bar 4}}

\newcommand{\ah}{\hat{a}}
\newcommand{\bh}{\hat{b}}
\newcommand{\ch}{\hat{c}}

\newcommand{\jb}{\bar{\jmath}}
\newcommand{\kb}{\bar{k}}

\newcommand{\Cyl}{{\rm Cyl}}
\newcommand{\cyl}{{\rm cyl}}

\newcommand{\mfrak}{{\mathfrak{m}}}
\newcommand{\gfrak}{{\mathfrak{g}}}
\newcommand{\hfrak}{{\mathfrak{h}}}

\newcommand{\C}{\mathbb C}
\newcommand{\R}{\mathbb R}

\newcommand{\Acal}{{\cal A}}

\newcommand{\Hcal}{{\cal H}}
\newcommand{\Fcal}{{\cal F}}
\newcommand{\Ecal}{{\cal E}}

\newcommand{\Pcal}{{\cal P}}

\newcommand{\Mcal}{{\cal M}}

\def\im{{\rm i}}
\def\e{{\,\rm e}\,}
\def\N2{$N{=}2$}
\def\pa{\partial}
\def\diff{{\rm d}}

\def\sfrac#1#2{{\textstyle\frac{#1}{#2}}}
\def\>{\rangle}
\def\<{\langle}
\def\+{\dagger}
\def\={\ =\ }

\def\and{\qquad\textrm{and}\qquad}
\def\for{\qquad\textrm{for}\quad}
\def\with{\qquad\textrm{with}\quad}

\def\ph{\phantom{-}}

\begin{document}
\begin{titlepage}
\setcounter{page}{0}
\begin{flushright}
ITP--UH--08/14
\end{flushright}

\vskip 2.5cm

\begin{center}

{\LARGE\bf Instantons on sine-cones over Sasakian manifolds
}

\vspace{15mm}

{\Large 
Severin Bunk${}^\+$,\
Tatiana A. Ivanova${}^*$, \
Olaf Lechtenfeld${}^{\+\times}$,}\\[8pt]
{\Large 
Alexander D. Popov${}^\+$} \ and \  {\Large Marcus Sperling${}^\+$}\\[10mm]
\noindent
${}^\+${\em Institut f\"ur Theoretische Physik} \\
{\em Leibniz Universit\"at Hannover}\\
{\em Appelstra\ss e 2, 30167 Hannover, Germany}\\[2mm]
Email: {\tt Severin.Bunk, Olaf.Lechtenfeld, Alexander.Popov, Marcus.Sperling@itp.uni-hannover.de}\\[5mm]
${}^\times${\em Riemann Center for Geometry and Physics\\
Leibniz Universit\"at Hannover \\
Welfengarten 1, 30167 Hannover, Germany }\\[5mm]
\noindent ${}^*${\em Bogoliubov Laboratory of Theoretical Physics, JINR} \\
{\em  141980 Dubna, Moscow Region, Russia}\\[2mm]
Email: {\tt ita@theor.jinr.ru}

\noindent

\vspace{15mm}

\begin{abstract}
\noindent
We investigate instantons on sine-cones over Sasaki-Einstein and 3-Sasakian manifolds. 
It is shown that these conical Einstein manifolds are K\"ahler with torsion (KT) manifolds 
admitting Hermitian connections with totally antisymmetric torsion.
Furthermore, a deformation of the metric on the sine-cone over 3-Sasakian manifolds 
allows one to introduce a hyper-K\"ahler with torsion (HKT) structure. 
In the large-volume limit these KT and HKT spaces become Calabi-Yau and hyper-K\"ahler conifolds, 
respectively. We construct gauge connections on complex vector bundles over conical KT and 
HKT manifolds which solve the instanton equations for Yang-Mills fields in higher dimensions.
\end{abstract}

\end{center}

\end{titlepage}

\section{Introduction}

\noindent
K\"ahler geometry is by now an established mathematical field with strong interrelations to theoretical physics, especially to supersymmetry and string theory.
Another important geometry used in string theories and M-theory is Sasakian geometry (see~\cite{1}-\cite{3} for reviews and references), 
especially in the AdS/CFT correspondence in the physically interesting dimensions five and seven.
Here, we shall focus on Sasaki-Einstein manifolds and on 3-Sasakian manifolds, which are of dimension $2n{+}1$ and $4n{+}3$, respectively.

On a manifold $M$ of real dimension $2n{+}1$, Sasakian geometry is sandwiched between K\"ahlerian geometries on particular manifolds of the two neighboring
dimensions. In particular, the metric cone over any Sasakian manifold is K\"ahler, and over any 3-Sasakian manifold it is hyper-K\"ahler.
If the metric on a Sasakian manifold is Einstein, then the metric cone over it is a Calabi-Yau manifold.
As examples, some well-known homogeneous and some recently discovered inhomogeneous Sasakian spaces often occur in string compactifications~\cite{4}-\cite{6}. 
These manifolds admit a connection with non-vanishing torsion and a structure group of
SU($n)\subset\ $SO$(2n{+}1)$ or Sp($n)\subset\ $SO$(4n{+}3)$,
respectively. There exist brane solutions of ten-dimensional supergravity which interpolate between an Ad$S_{p+1}\times X_{9-p}$ near-horizon geometry and an asymptotic
geometry $M_p\times C(X_{9-p})$, where $M_p$ is $p$-dimensional Minkowski  space and $C(X_{9-p})$ is a metric cone over $X_{9-p}$ (see e.g.~\cite{7,8}
and references therein). Such kinds of brane solutions in heterotic supergravity with Yang-Mills instantons on the metric cones $C(X_{9-p})$ were considered in~\cite{9,10}.
We intend to generalize them by considering sine-cones with a K\"ahler-torsion structure instead of metric cones with a K\"ahler structure.

In this paper we will consider sine-cones over Sasaki-Einstein and 3-Sasakian manifolds and solve instanton equations for  Yang-Mills fields on these conical manifolds.
Recall that for any Riemannian metric $g$ on a manifold $M$, the warped product metric $\gt$ on $C(M)=\R_+\times M$ is defined as
\beq
\gt \= \diff r^2 + f^2(r)\, g\ ,
\eeq
where $r\in\R_+$ and $f(r)$ is a {\it warping\/}  function. 
$(C(M), \gt)$ is called the {\it metric cone\/} over $M$ if $f(r)=r$, and it is known as the {\it sine-cone\/} over $M$ if $f(r)=\sin r$.
It is known that the sine-cone metric $\gt$ over a Sasaki-Einstein or 3-Sasakian manifold is Einstein~\cite{2}.\footnote{The
corresponding metric-cone metric is even Ricci-flat.} We show that these conical manifolds admit K\"ahler with torsion (KT) structures, i.e.\ on them there exists a Hermitian connection with a
totally antisymmetric torsion $\tilde T= J\,\diff\ot$. Here, $J$ is an almost complex structure, and $\ot (\cdot ,\cdot )=\gt (J\,\cdot,\cdot)$ is the fundamental 2-form. For $\tilde T=0$
 the Hermitian structure is K\"ahler.

We also show that for any 3-Sasakian manifold one can deform the sine-cone metric $\gt$ in such a way that $C(M)$ will be a hyper-K\"ahler with torsion (HKT) manifold.
HKT geometry has been described in detail e.g.\ in~\cite{11} and intensively studied since then (see e.g.~\cite{12}-\cite{14} and references therein). In fact, these Hermitian manifolds
with three integrable almost complex structures $J^\a$, $\a =1,2,3$, are not hyper-K\"ahler for non-vanishing torsion $\tilde T= J^1\diff\ot^1=J^2\diff\ot^2=J^3\diff\ot^3$, but we
will follow the established terminology. Here, $\ot^\a (\cdot ,\cdot )=\gt (J^\a\,\cdot,\cdot)$ are three Hermitian 2-forms.

After introducing the KT and HKT structures on the sine-cones over Sasaki-Einstein and 3-Sasakian manifolds, we consider the appropriate instanton equations on these conical manifolds.
Such first-order gauge equations, which generalize the Yang-Mills anti-self-duality equations in $d{=}4$ to higher-dimensional manifolds with special holonomy (or, more generally, $G$-structure),
were introduced and studied both in the physical~\cite{15}-\cite{17} and mathematical~\cite{18}-\cite{20} literature.
Some instanton solutions were found e.g.\ in~\cite{21}-\cite{27}.
Due to the foliated structure of the considered conical KT and HKT manifolds, a natural ansatz
for the gauge fields reduces the instanton equations to matrix-model equations. 
We discuss their simplest analytic and numerical solutions.

This article is arranged as follows. In Section~2 we collect various geometric facts concerning Sasaki-Einstein, 3-Sasakian, K\"ahler-torsion and hyper-K\"ahler with torsion manifolds. We introduce
the KT and HKT structures on sine-cones over Sasaki-Einstein and 3-Sasakian manifolds. In Section~3 we discuss the instanton equations in more than four dimensions and specialize them for KT
and HKT manifolds. Then we describe an ansatz reducing these instanton equations to matrix equations and present some solutions.

\bigskip

\section{Hermitian manifolds with torsion}

\noindent
{\bf KT manifolds.} A $G$-structure on a smooth orientable manifold $M$ of dimension $m$ is a reduction of the structure group GL$(m,\R)$ of the tangent bundle $TM$
to a closed subgroup $G\subset\ $GL$(m, \R)$.  Choosing an orientation and a Riemannian metric $g$ defines an  SO$(m)$-structure on $M$.  We assume that $m=2n$ is even and that
$(M, g)$ is an almost Hermitian manifold. This means that there exists an almost complex structure $J\in \mbox{End}(TM)$, with $J^2=-{\bf {1}}_{TM}$, which is compatible with the metric $g$,
i.e.\ $g(JX,JY)=g(X,Y)$ for all $X,Y\in TM$. One can introduce the fundamental two-form $\ome$ as
\beq\label{2.1}
\ome (X,Y)\ :=\ g(JX, Y)\for X,Y\in TM\ ,
\eeq
and the canonical objects $(g, J, \ome )$ define a U$(n)$-structure on $M$. The additional existence of a complex $n$-form $\Om$ reduces the U$(n)$- to an SU$(n)$-structure. It implies
that the almost Hermitian manifold $(M, g, J)$ has a topologically trivial canonical bundle. If the almost complex structure $J$ is integrable, then $(M, g, J)$ is a Hermitian manifold and $(M, g, J, \Om)$ is Calabi-Yau.

On any almost Hermitian manifold $(M, g, J)$ one has a Hermitian connection\footnote{This connection $\nabla = \diff + \Gamma$ preserves $g, J$ and $\ome$.} with totally antisymmetric
torsion $T$. Two particularly interesting cases arise when the torsion $T$ is the real part of either a (3,0)-form or a (2,1)-form. In the former case, the manifold $M$ is called nearly K\"ahler~\cite{28,29}.
The latter case entails a property called K\"ahler-torsion (KT)~\cite{30,31}. Namely, on a Hermitian manifold  $(M, g, J)$ the KT connection is a Hermitian connection $\Gamma$ with antisymmetric
torsion $T$ given by
\beq\label{2.2}
T=J\diff\ome\ ,
\eeq
where $J$ acts on the $p$-form $e^{a_1}\wedge \cdots \wedge e^{a_p}$ as
\beq\label{2.3}
J\left(e^{a_1}\wedge e^{a_2}\wedge \cdots \wedge e^{a_p}\right) \= Je^{a_1}\wedge Je^{a_2}\wedge \cdots \wedge Je^{a_p}\with
Je^{a}=J^{a}_{b}e^{b}\ .
\eeq
Here $\{e^a\}$ with $a=1,\ldots,2n$ is a local frame for the cotangent bundle $T^*M$, and $J^a_b$ are the corresponding components of the
almost complex structure $J$. Note that $J$ is integrable in the KT case and non-integrable in the nearly K\"ahler case.
For traceless anti-Hermitian $\Gamma$ the KT manifolds are called Calabi-Yau torsion~\cite{32}.

\noindent {\bf Remark.} Let
\beq\label{2.4} \t^j:= e^{2j-1} + \im e^{2j}\and
\t^{\jb}:=\overline{\t^j}\with j=1,\ldots,n \eeq
constitute a local frame for the (1,0) and (0,1)
parts of the complexified cotangent bundle. Then $J$, $g$ and $\ome$ may be chosen as
follows:
\beq\label{2.5} J\t^j = \im \t^j\ ,\qquad g=\sum_{j=1}^n\t^j{\otimes}\,\t^{\jb}\and \ome
=\sfrac{\im}2\sum_{j=1}^n\t^j\wedge\t^{\jb}\ . \eeq
Their non-vanishing components are thus given by
\beq\label{2.6} J^{~j}_k=\im\de^{~j}_k\ ,\quad
J^{\jb}_{~\kb}=-\im\de^{\jb}_{~\kb}\ ,\quad g_{j\kb}=\sfrac12\de_{j{\kb}}\and
\ome_{j{\kb}}={\sfrac{\im}2}\de_{j{\kb}} \eeq
with respect to the $\t$-basis. For
$g^{-1}$ and $\ome^{-1}$ we have
\beq\label{2.7} g^{j\kb}=2\de^{j{\kb}}\and
\ome^{j{\kb}}=-2{\im}\de^{j{\kb}}\ . \eeq

\noindent 
{\bf Cones.} For any Riemannian manifold $(\Mcal , g)$ we define
\beq\label{2.8} C_{\L}(\Mcal) \=\bigl( (0, \L\pi)\times\Mcal ,\ \gt\bigr) \with\quad
\gt \=\diff r^2 + \L^2\sin^2(\sfrac{r}{\L})\,g \eeq
to be the {\it sine-cone\/} over $\Mcal$. Here $r\in (0, \L\pi)$ which is an open interval, and so the volume of the
sine-cone is $\L\,$vol$\Mcal$. In the infinite-volume limit $\L\to\infty$ the sine-cone becomes the {\it metric 
cone\/}\footnote{One usually omits the adjective ``metric'' and simply says ``cone''.}
\beq\label{2.9} C_{\infty}(\Mcal)\equiv C(\Mcal)\= \bigl(\R_+\times\Mcal,\ \gt\bigr) \with\quad \gt \=\diff r^2 + r^2g\ . \eeq
Note that
\beq\label{2.10} \gt \=\diff r^2 + \L^2\sin^2(\sfrac{r}{\L})\  g\ =:\ \L^2\sin^2\!\vp\ (\diff\tau^2+g)\ , \eeq where
\beq\label{2.11}
\vp=\sfrac{r}{\L}\and \tau = \log\bigl(2\sfrac{\L}{r_0} \tan\sfrac{\vp}{2}\bigr)\quad\Leftrightarrow\quad
r=2\L\arctan\Bigl(\frac{r_0\e^{\tau}}{2\L}\Bigr) \eeq
with $\tau\in\R$ and a constant $r_0\in\R_+$. In the limit $\L\to\infty$, (\ref{2.11}) simplifies to
\beq\label{2.12} \tau =\log(\sfrac{r}{r_0}) \quad\Leftrightarrow\quad r=r_0 \e^{\tau} \eeq 
which are valid for the metric cone with
\beq\label{2.13} \gt\=\diff r^2 + r^2g\=
r^2\Bigl(\frac{\diff r^2}{r^2} + g\Bigr)\= r_0^2\e^{2\tau}(\diff\tau^2+g)\ .
\eeq 
It follows from (\ref{2.12}) and (\ref{2.13}) that both cones are
conformally equivalent to the cylinder 
\beq\label{2.14} \Cyl(\Mcal)\=\bigl(\R\times\Mcal ,\ g_{\cyl}\bigr) 
\with\quad g_{\cyl}\=\diff\tau^2 +g\ .\eeq 

\medskip

\noindent
{\bf Sasaki-Einstein manifolds.} A $(2n{+}1)$-dimensional Riemannian manifold $(\Mcal , g)$ is called {\it Sasakian\/} if the metric cone $(C(\Mcal ), \gt )$ is
K\"ahler.\footnote{This is one of several equivalent definitions of Sasakian manifolds \cite{2}.} A Sasakian manifold $(\Mcal , g)$ is
{\it Sasaki-Einstein\/} if in addition the metric $g$ is Einstein. In this case the metric cone $C(\Mcal)$ is a Ricci-flat  K\"ahler manifold (Calabi-Yau), so its holonomy group
is reduced from U$(n)$ to SU$(n)$. Sasaki-Einstein manifolds have
a reduced structure group of SU$(n)\subset\ $SO$(2n{+}1)$. They are endowed with 1-, 2-, 3- and 4-forms $\h , \ome , P$ and $Q$, which can be defined in a local orthonormal basis
$\{e^{\ah}\}$, $\ah = 1,\ldots, 2n{+}1$, as
\beq\label{2.16}
\h = -e^{2n{+}1}\ ,\quad \ome  =\sum_{j=1}^n e^{2j-1}\wedge e^{2j}\ ,\quad P=\h\wedge\ome\and Q=\sfrac12\,\ome\wedge\ome\ .
\eeq
These forms satisfy the relations
\beq\label{2.17}
\diff e^{2n{+}1} = -2\ome\and \diff P = 4Q\ .
\eeq
Note that, in the above basis, the torsion $T$ of the canonical $su(n)$-valued connection 
$\Gamma$ on $\Mcal$ is not totally antisymmetric and has the components (see e.g.~\cite{9})
\beq\label{2.18}
T^a \= \sfrac{n{+}1}{2n}\, P_{a\bh\ch}\, e^{\bh}\wedge e^{\ch}\and T^{2n{+}1} \= P_{2n{+}1\,\bh\ch}\, e^{\bh}\wedge e^{\ch}\ ,
\eeq
where $\{\ah\} = \{a, 2n{+}1\}$ with $a=1,\ldots,2n$.

\bigskip

\noindent
{\bf 3-Sasakian manifolds.} A $(4n{+}3)$-dimensional Riemannian manifold $(\Mcal , g)$ is called  
{\it 3-Sasakian\/} if the metric cone  $(C(\Mcal ), \gt )$ on $\Mcal$
is hyper-K\"ahler. The structure group of $\Mcal$ then is Sp$(n)\subset\ $SO$(4n{+}3)$, 
and we let the index $a$ run from $1$ to~$4n$. Note that any 3-Sasakian manifold is Einstein and can be endowed with three 1-forms $\h^\a$, three 2-forms
$\ome^\a$, a 3-form $P$ and a 4-form $Q$ with $\a=1,2,3$~\cite{1,2}. In a local orthonormal co-frame $\{e^{\ah}\}$ where $\{\ah\} = \{a, 4n {+}\a\}$, these forms can be written as
\bea\label{2.19}
\h^{\a}&=& - e^{4n{+}\a} \ ,\vphantom{\Big|}
\\ \label{2.20}
\ome^1&=&\sum_{j=1}^n \left(e^{4j-3}\wedge e^{4j} + e^{4j-2}\wedge e^{4j-1}\right)\ ,
\\ \label{2.21}
\ome^2&=&\sum_{j=1}^n \left(-e^{4j-3}\wedge e^{4j-1} + e^{4j-2}\wedge e^{4j}\right)\ ,
\\ \label{2.22}
\ome^3&=&\sum_{j=1}^n \left(e^{4j-3}\wedge e^{4j-2} + e^{4j-1}\wedge e^{4j}\right)\ ,
\\ \label{2.23}
P&=&\sfrac13\, \bigl({\textstyle\sum_\a}\,\h^{\a}\wedge\ome^{\a} + \eta^{123}\bigr)
\= -\sfrac13\,\bigl({\textstyle\sum_\a}\,\ome^{\a}\wedge e^{4n{+}\a }+e^{4n{+}1}\wedge e^{4n{+}2}\wedge e^{4n{+}3}\bigr) \ ,\vphantom{\bigg|}
\\ \label{2.24}
Q&=&\sfrac16\, {\textstyle\sum_\a}\,\ome^{\a}\wedge\ome^{\a} \ .\vphantom{\bigg|}
\eea
The forms $e^{4n{+}\a}$ and $\ome^\a$ satisfy the differential identities
\beq\label{2.25}
\diff e^{4n{+}\a }\=-\ve^{\a}_{\b\g} e^{4n{+}\b }\wedge  e^{4n{+}\g } - 2\ome^{\a}\ ,
\eeq
\beq\label{2.26}
\diff \ome^{\a }\=-2\ve^{\a}_{\b\g} e^{4n{+}\b }\wedge \ome^{\g}\ ,
\eeq
where $\ve$ is the Levi-Civita tensor.
The torsion $T$ of the canonical $sp(n)$-valued connection $\Gamma$ on any 3-Sasakian manifold takes the form (see e.g.~\cite{9})
\beq\label{2.27}
T^a \= \sfrac{3}{2}\, P_{a\bh\ch}\, e^{\bh}\wedge e^{\ch}\and 
T^{\a} \= 3P_{4n{+}\a\,\bh\ch}\, e^{\bh}\wedge e^{\ch}\ .
\eeq
$T$ is not totally antisymmetric for the Einstein metric on $\Mcal$.

\bigskip

\noindent
{\bf KT structure on sine-cones.}  A well-known theorem~\cite{2} states that, if $(\Mcal , g)$ is a $k$-dimensional Einstein
manifold with Einstein constant $k{-}1$, then the sine-cone  $(C(\Mcal ), \gt )$ over $\Mcal$ with the metric (\ref{2.10}) for $\L=1$ is Einstein
with  Einstein constant $k$. Here we will show that the sine-cone over any Sasaki-Einstein manifold is not only  Einstein
but also carries a K\"ahler-torsion structure.

Consider the cylinder 
\beq\label{2.29} 
\Cyl(\Mcal)\=\bigl(\R\times\Mcal ,\ g_{\cyl}\bigr) 
\with\quad g_{\cyl} \= \de_{\ah\bh} e^{\ah}{\otimes}\,e^{\bh} + e^{2n+2}{\otimes}\,e^{2n+2}\ , \eeq
where $\{\ah\} = \{a, 2n{+}1\}$ with $a=1,\ldots,2n$, 
and compare it to the sine-cone
\beq\label{2.28}
M^{2n+2}\ :=\ C_{\L}(\Mcal) \= \bigl( (0, \L\pi)\times\Mcal ,\ \gt\bigr)
\eeq
parametrized via
\beq\label{2.30} e^{2n+2}=\diff\tau =\frac{\diff\vp}{\sin\vp}\ ,\quad \vp=\frac{r}{\L}\
,\quad \tau\in\R \ ,\quad \vp\in (0,\pi )\ . \eeq 
Then the local basis $\{\et^{\ah},\et^{2n+2}\}$ on the latter is defined as
\beq\label{2.31}
\et^{\ah} = \L\sin\vp\  e^{\ah}\and \et^{2n+2}= \L\sin\vp\  e^{2n+2}= \diff r \ ,
\eeq 
and its metric reads
\beq\label{2.32} \gt \= \de_{\ah\bh}
\et^{\ah}{\otimes}\,\et^{\bh} + \et^{2n+2}{\otimes}\,\et^{2n+2}\ . \eeq

Let us also introduce the 2-form
\beq\label{2.33}
\ot\ :=\ \L^2 \sin^2\vp\ (\ome + e^{2n+1}\wedge e^{2n+2})\ ,
\eeq
where $\ome$ is the 2-form defined in (\ref{2.16}) and obeying (\ref{2.17}). It is easy to check that
\beq\label{2.34}
\diff\ot \= \frac{2}{\L} \frac{\cos\vp -1}{\sin\vp}\ \ot\wedge\et^{2n+2} 
\= - \frac{2}{\L} \tan\sfrac{\vp}{2}\ \ot\wedge \et^{2n+2}\ .
\eeq
The canonical almost complex structure $J$ on $M^{2n+2}$ is fixed by $\gt$ and $\ot$ via (\ref{2.4})--(\ref{2.6})
but with the range of $j$ extended to $n{+}1$.

Finally one can define the torsion
\beq\label{2.35}
\tilde T\ :=\ J\diff\ot \=  - \frac{2}{\L} \tan\sfrac{\vp}{2}\ \ot\wedge \et^{2n+1}
\eeq
which is proportional to $P$ from (\ref{2.16}).
Since $\tilde T$ is of type (2,1)+(1,2) w.r.t.~$J$, the almost complex structure $J$ is integrable, 
and we obtain a KT structure on the sine-cone (\ref{2.28}) over any Sasaki-Einstein manifold.

\bigskip

\noindent {\bf Conical HKT structure.} The sine-cone 
\beq
M^{4n+4} \= C_\L(\Mcal) \= \bigl( (0, \L\pi)\times\Mcal ,\ \gt\bigr)
\eeq
over a $(4n{+}3)$-dimensional 3-Sasakian manifold $(\Mcal, g)$ with a metric (\ref{2.10}) is an Einstein
manifold since $(\Mcal, g)$ is Einstein. 
Since $\bigl(\Mcal,e^{4n+3},\ome^3{+}e^{4n+1}{\wedge}e^{4n+2},g\bigr)$ is Sasaki-Einstein,
the previous subsection applies, and one can introduce a KT structure on $M^{4n+4}$
by choosing the 2-form
\beq\label{2.36} \ot\ :=\ \L^2 \sin^2\!\vp\ (\ome^3 + e^{4n+1}\wedge e^{4n+2}+
e^{4n+3}\wedge e^{4n+4}) \eeq 
as in (\ref{2.33})--(\ref{2.35}),
where $\ome^3$ is defined by (\ref{2.22}) and
\beq\label{2.37} e^{4n+4} = \diff\tau\ ,\quad \tau = \log\bigl(2\sfrac{\L}{r_0}\,
\tan\sfrac{\vp}2\bigr)\ ,\quad \vp=\sfrac{r}{\L}\ . \eeq
Here $\{e^{\ah}, e^{4n+4}\}=\{e^a, e^{4n+\a}, e^{4n+4}\}$ with $a=1,\ldots,4n$ 
is a local basis of one-forms on the cylinder
$\Mcal\times\R$ with the metric
\beq\label{2.38} g^{}_{\cyl}\=\de_{ab}e^a{\otimes}\,e^b +
\de_{\m\n}e^{4n+\m}{\otimes}\,e^{4n+\n}\ , \eeq
where we introduced the index set $\{\mu\} = \{\a,4\}$.
Recall that $e^{\ah}$ and $\ome^\a$, defined in (\ref{2.19})--(\ref{2.22}), satisfy the
identities (\ref{2.25}) and~(\ref{2.26}).

We have
\beq\label{2.39}
\ome^{\a}\=\sfrac12\, \ome^{\a}_{ab}e^a\wedge e^b\for a=1,\ldots,4n\ ,
\eeq
where the components $\ome^{\a}_{ab}$ of the 2-forms $\ome^{\a}$ can be read off from  (\ref{2.20})--(\ref{2.22}). 
For later use we define three more 2-forms,
\beq\label{2.40}
\ome^\a_{\bot}\ :=\ \sfrac12\,\h^{\a}_{\m\n} e^{4n+\m}\wedge  e^{4n+\n}\ ,
\eeq
where $\h^{\a}_{\m\n}$ are the components of the 't~Hooft tensor,
\beq\label{2.41}
\h^{\a}_{\b\g}=\ve^{\a}_{\b\g} \and  \h^{\a}_{\b 4}=-\h^{\a}_{4\b}=\de^{\a}_{\b}\ .
\eeq
Using  (\ref{2.39})--(\ref{2.41}), we may introduce on $M^{4n+4}$ three almost complex structures $J^{\a}$
with components
\beq\label{2.43}
J^{\a a}_{~\,b}=\ome^{\a}_{bc}\de^{ca}\and J^{\a\, 4n+\m}_{~\,4n+\n}=\h^{\a}_{\n\s}\de^{\s\m}\ .
\eeq
It is not difficult to show that
\beq\label{2.44}
J^{\a}J^{\b} \= -\de^{\a\b}\textrm{id} + \ve^{\a\b\g}J^{\g}\ ,
\eeq
i.e.\ the three almost complex structures define a quaternionic structure on $M^{4n+4}$.

The 2-form $\ot$ in~(\ref{2.36}) is of type (1,1) w.r.t.~$J^3$. Differentiating, we find
\beq\label{2.45}
\tilde T\=  J^3\diff\ot \=  - \sfrac{2}{\L} \tan\sfrac{\vp}{2}\ \ot\wedge \et^{4n+3}\ ,
\eeq
where we use local co-frame fields
\beq\label{2.46}
\et^a=\L\sin\vp\, e^a\and \et^{4n+\m}=\L\sin\vp\,e^{4n+\m}
\eeq
on $M^{4n+4}=C_\L(\Mcal)$. Thus, $M^{4n+4}$ allows for a KT structure.

In order to extend this to a HKT structure on  $M^{4n+4}$, we need three 2-forms
\beq\label{2.47}
\ot^{\a}\=\L^2\sin^2\!\vp\ (f_1\ \ome^{\a}+f_2\ \ome^{\a}_{\bot})\ ,
\eeq
where $f_1=f_1(\vp)$ and  $f_2=f_2(\vp)$ are yet undefined real functions, 
and $\ome^{\a}$ and $\ome^{\a}_{\bot}$ are given by (\ref{2.39}) and (\ref{2.40}), respectively.
Taking the exterior derivative of (\ref{2.47}), we obtain
\beq\label{2.48}
\diff \ot^{\a}\=\L^2\sin^2\!\vp\,\Bigl\{ \bigl(B_1\,\ome^{\a} + \sfrac12\,B_2\,\ve^{\a}_{\b\g}e^{4n+\b}\wedge e^{4n+\g}\bigr)\wedge e^{4n+4} + 
B_3\,\ve^{\a}_{\b\g}e^{4n+\b}\wedge \ome^{\g}\Bigr\}\ ,
\eeq
where
\bea\label{2.49}
B_1 &=&\dot f_1\sin\vp + 2 f_1\cos\vp -2f_2\ ,
\\[3pt] \label{2.50}
B_2 &=&\dot f_2\sin\vp + 2 f_2\cos\vp -2f_2\ ,
\\[4pt] \label{2.51}
B_3 &=& -2(f_1 -f_2)\ ,
\eea
and the overdot indicates a derivative w.r.t.~$\vp$.
The definitions of $J^\a$ imply that
\beq\label{2.52}
J^\a\ome^\b \= (-1)^{1+\de_{\a\b}}\ome^\b  \and
\eeq
\beq\nonumber
J^1e^{4n+1}=-e^{4n+4}\ ,\quad J^2e^{4n+1}=\ph e^{4n+3}\ ,\quad J^3e^{4n+1}=-e^{4n+2}\ ,
\eeq
\beq\label{2.53}
J^1e^{4n+2}=-e^{4n+3}\ ,\quad J^2e^{4n+2}=-e^{4n+4}\ ,\quad J^3e^{4n+2}=\ph e^{4n+1}\ ,
\eeq
\beq\nonumber
J^1e^{4n+3}=\ph e^{4n+2}\ ,\quad J^2e^{4n+3}=-e^{4n+1}\ ,\quad J^3e^{4n+3}=-e^{4n+4}\ ,
\eeq
\beq\nonumber
J^{\a}e^{4n+4}=\ph e^{4n+\a}\ .
\eeq

In \cite{13a} it has been proven that the HKT condition is equivalent to
\beq\label{2.54}
J^1\diff\ot^1 \= J^2\diff\ot^2 \= J^3\diff\ot^3 \= \tilde T \ ,
\eeq
where $\tilde T$ is the torsion of the $sp(n{+}1)$-valued hyper-Hermitian connection on $M^{4n+4}$.
Using  (\ref{2.52}) and (\ref{2.53}) as well as demanding that $\tilde T$ is proportional to $P$ from (\ref{2.23}), 
we obtain from  (\ref{2.54}) the constraints
\beq\label{2.55}
B_1\= B_2\= B_3
\eeq
which are equivalent to the differential equations
\beq\label{2.56}
\dot f \sin\vp + 2f\cos\vp =0 \and \dot f_2 \sin\vp + 2f_2(\cos\vp -1) +2f =0 
\qquad\textrm{for}\qquad f:=f_1-f_2\ .
\eeq
Solutions of these equations can be chosen in the form
\beq\label{2.57}
f\=\frac{c}{\sin^2\vp} \and f_2\=\frac{2c_2}{\cos^4\sfrac{\vp}2} + \frac{c}{\sin^2\vp}
\qquad\Rightarrow\qquad f_1\=\frac{2c_2}{\cos^4\sfrac{\vp}2} + \frac{2c}{\sin^2\vp}\ ,
\eeq
where $c$ and $c_2$ are yet arbitrary constants of integration. In the limit $\L\to\infty$ 
we would like our HKT space $M^{4n+4}$ to coincide with
the standard hyper-K\"ahler metric cone $C(\Mcal)$ with vanishing torsion $\tilde T=0$.
This is achieved for
\beq\label{2.58}
c=\frac{c_1}{\L^3}\ ,
\eeq
where $c_1$ is constant. Then for $\L\to\infty$ we get
\beq\label{2.59}
\ot^\a\ \to\ 2c_2\,r^2(\ome^\a +\ome^\a_{\bot})\ =:\ \oh^\a \with \diff\oh^\a=0\ .
\eeq

The metric on the HKT manifold $M^{4n+4}$ with three Hermitian\footnote{Note that the conditions (\ref{2.54}) 
imply the integrability of the almost complex structures (\ref{2.43}).} 
structures (\ref{2.47}) takes the form
\beq\label{2.60}
\gt \= f_1\,\de_{ab}\et^a{\otimes}\,\et^b + f_2\,\de_{\m\n}\et^{\m}{\otimes}\,\et^{\n} 
\= f_2\bigl(\sfrac{f_1}{f_2}\,\de_{ab}\et^a{\otimes}\,\et^b + \de_{\m\n}\et^{\m}{\otimes}\,\et^{\n}\bigr)\ .
\eeq
It is conformally equivalent, with conformal factor $f_2$, to the metric on the sine-cone  $C_\L(\Mcal)$ over a 3-Sasakian manifold $\Mcal$.
The simplest case occurs for the choice $c_1=0$ when  $f_1=f_2$.

\bigskip

\section{Instantons on conical KT and HKT manifolds}

\noindent
{\bf Instanton equations.} Let $\Sigma$ be a differential form of degree $m{-}4$ on an $m$-dimensional Riemannian manifold $M$, and let $\Ecal$ be a complex vector
bundle over $M$ endowed with a connection $\Acal$. The  $\Sigma$-anti-self-duality (or instanton) equations are defined as the first-order equations~\cite{20}
\beq\label{3.1}
\ast \Fcal \= -\Sigma\wedge\Fcal
\eeq
for the connection $\Acal$ with curvature $\Fcal =\diff\Acal + \Acal\wedge\Acal$. Here $\ast$ is the Hodge duality operator on $M$. 
Taking the exterior derivative of (\ref{3.1}) and using the Bianchi identity, we obtain
\beq\label{3.2}
\diff\ast\Fcal\ +\ \Acal\wedge\ast\Fcal - (-1)^m\ast\Fcal\wedge\Acal\ +\ \ast\Hcal\wedge\Fcal \=0\ ,
\eeq
where the 3-form $\Hcal$ is defined by 
\beq\label{3.3}
\ast\Hcal\ :=\ \diff\Sigma\ .
\eeq

The second-order equations (\ref{3.2}) differ from the standard Yang-Mills equations by the last term involving a 3-form $\Hcal$, which can be identified
with a totally antisymmetric torsion on $M$. This torsion term naturally appears in string-theory compactifications with fluxes~\cite{34}. 
For $\diff\Sigma=0$, the torsion term vanishes and the instanton equations (\ref{3.1}) imply the ordinary Yang-Mills equations. 
The latter also holds true when the instanton solution $\Fcal$ satisfies $\diff\Sigma\wedge\Fcal=0$ as well, 
as e.g.\ on nearly K\"ahler 6-manifolds, nearly parallel $G_2$-manifolds and Sasakian manifolds~\cite{9}.

The torsionful Yang-Mills equations (\ref{3.2}) are the variational equations for the action 
\beq\label{3.4} 
S\=-\int_M \mbox{Tr}\bigl(\Fcal\wedge\ast\Fcal\ +\ \Fcal\wedge\Fcal\wedge\Sigma \bigr) \ ,
\eeq 
and the instanton equations (\ref{3.1}) can be derived from this action using a Bogomolny argument. In the case of a closed
form $\Sigma$, the second term in (\ref{3.4}) is topological and the torsion (\ref{3.3}) disappears from (\ref{3.2}).

If a manifold is endowed with a 4-form $Q$, a natural choice for the $(m{-}4)$-form $\Sigma$ in the
instanton equations (\ref{3.1}) will be its Hodge dual, $\Sigma\propto*Q$. Therefore,
on KT manifolds with $m=2n{+}2$ one should take
\beq\label{3.5}
\tilde{Q}_{\rm KT}\=\sfrac12\,\ot\wedge\ot \qquad\textrm{in}\qquad
\ast \Fcal \= -\ast\tilde{Q}_{\rm KT}\wedge\Fcal\ .
\eeq
On HKT manifolds there exist three 2-forms $\ot^\a$, from which one can build the 4-form
\beq\label{3.6}
\tilde{Q}_{\rm HKT}\=\sfrac16\,\ot^\a\wedge\ot^\a \qquad\textrm{in}\qquad
\ast\Fcal \= -\ast\tilde{Q}_{\rm HKT}\wedge\Fcal\ .
\eeq

\noindent
{\bf Reduction to matrix equations.} Recall that the instanton equations on the cone $C(\Mcal )$ over Sasaki-Einstein or 3-Sasakian manifolds $\Mcal$
are equivalent to the equations on the cylinder $\Cyl(\Mcal)$~\cite{9} with the metric
\beq\label{3.7}
\gt \= \diff\tau^2 + g\ ,
\eeq
where $g$ is the metric on $\Mcal$ and $\tau$ is related with $r$ by (\ref{2.11}). Let us denote by $G$ and $H$ the structure groups of the canonical connection on
$C(\Mcal)$ and $\Mcal$, respectively. In the KT case we have $(G,H)=($SU$(n{+}1)$, SU$(n)$), and in the HKT case $(G, H)=($Sp$(n{+}1)$, Sp$(n)$). 
As a vector space, the Lie algebra $\gfrak=\mbox{Lie}\,G$ decomposes into $\hfrak=\mbox{Lie}\,H$ 
and its orthogonal complement $\mfrak$,
\beq\label{3.8}
\gfrak \=\hfrak \oplus\mfrak\ .
\eeq 
The vector space $\mfrak$ can be identified with the linear span of the orthonormal basis $\{e^{\ah}\}$ on~$T^*\Mcal$.

In any given irreducible representation $\rho$ of $\gfrak$, 
the generators $I_i$ of $\hfrak$ and $I_{\ah}$ of $\mfrak$ obeying the commutation relations
\beq\label{3.9}
[I_i, I_j]\=f^k_{ij}\, I_k \ ,\quad [I_i, I_{\ah}]\=f^{\bh}_{i\ah}\, I_{\bh}
\and
[I_{\ah}, I_{\bh}]\=f^i_{\ah\bh}\, I_i + f^{\ch}_{\ah\bh}\, I_{\ch}
\eeq
act on a representation space $V\cong\C^N$, i.e.\ $\rho: \gfrak\to\mathrm{End}(V)$.  
Consider a complex vector bundle $\Ecal \to \Cyl(\Mcal)$ such that the fibres are copies of~$V$.
Since $H$ is a closed subgroup of $G$, it also acts on the fibres of $\Ecal$, but the restriction 
of our $\gfrak$-representation~$\rho$ to the subalgebra~$\hfrak$ in general decomposes into a direct sum
of several irreducible $\hfrak$-representations, with the corresponding invariant subspaces comprising~$V$.

The canonical connection $\Ga$ on $T\Mcal$ is always an instanton~\cite{9,10}.
On the bundle $\Ecal$, it induces the $\rho(\hfrak)$-valued connection (denoted by the same letter)
\beq\label{3.10}
\Ga := \Ga^i I_i\ .
\eeq
Its curvature
\beq\label{3.11}
 R \= \diff\Ga + \Ga\wedge\Ga \= \bigl(\diff\Ga^i + \sfrac12\, f^i_{jk}\Ga^j\wedge\Ga^k\bigr)\,I_i
\eeq
satisfies the instanton equations (\ref{3.1})~\cite{9,10}.

Let us consider some matrix-valued functions $X_{\ah}(\tau)\in$ End$(V)$ 
and introduce on $\Ecal$ a $\rho(\gfrak)$-valued connection
\beq\label{3.12}
\Acal\ :=\ \Ga + X_{\ah} e^{\ah}\ .
\eeq
For $X_{\ah}$ depending on all coordinates of $\Cyl(\Mcal)$, this is the general form of a connection on the bundle  $\Ecal \to \Cyl(\Mcal)$.
Below, we shall impose independence of $X_{\ah}$ on the coordinates of $\Mcal$ and certain equivariance conditions, which will reduce (\ref{3.1}) to ordinary differential equations for $X_{\ah}$.

Recall that
\beq\label{3.13}
\diff e^{\ah} \= - \Ga^{\ah}_{\bh}\wedge e^{\bh} + T^{\ah} \= -\Ga^i f^{\ah}_{i\bh}\wedge e^{\bh} +\sfrac12\,
T^{\ah}_{{\bh}\ch}e^{\bh}\wedge e^{\ch} \ ,
\eeq
where $f^{\ah}_{i\bh}$ are the structure constants from (\ref{3.9}). From (\ref{3.12}) and (\ref{3.13}) it follows that
\beq\label{3.14}
\Fcal \=
\diff\Acal+\Acal{\wedge}\Acal \= R+\sfrac12\bigl([X_{\ah} , X_{\bh}]+T^{\ch}_{{\ah}{\bh}}X_{\ch} \bigr)e^{\ah}{\wedge}e^{\bh}+
\dot X_{\ah}\,\diff\tau{\wedge}e^{\ah}+\Ga^i{\wedge}e^{\ah} \bigl([I_i , X_{\ah}]- f^{\bh}_{i{\ah}} X_{\bh} \bigr)\ ,
\eeq
where $R$ is given in (\ref{3.11}) and $\dot X_{\ah} \equiv\sfrac{\diff}{\diff\tau}X_{\ah}$. In \cite{26}  it was shown that $\Fcal$
solves the instanton equations (\ref{3.1}) if the following matrix equations hold:
\beq\label{3.15}
[I_i, X_{\ah} ]\= f^{\bh}_{i{\ah}}X_{\bh}\ ,\ph
\eeq
\beq\label{3.16}
[X_{\ah} , X_{\bh} ]\ +\ T^{\ch}_{{\ah}{\bh}}X_{\ch} \= N^{\ch}_{{\ah}{\bh}}\dot X_{\ch}\ +\ f^i_{{\ah}{\bh}}N_i(\tau )\ .
\eeq
Here $N^{\ch}_{{\ah}{\bh}}$ is some constant tensor which we shall specify below for each case, and $N_i$ are some $\rho(\hfrak)$-valued functions defined by
(\ref{3.16}) after resolving the algebraic constraints (\ref{3.15}) and substituting their solutions $X_{\ah}$ into (\ref{3.16}). 
For $X_{\ah}$ satisfying (\ref{3.15}) and (\ref{3.16}), we have
\beq\label{3.17}
\Fcal\= R\ +\ \sfrac12\,N_if^i_{\ah\bh}\,e^{\ah}\wedge e^{\bh}\ +\
\dot X_{\ah} \bigl(\diff\tau\wedge e^{\ah} + \sfrac12\,N^{\ah}_{\bh\ch}\,e^{\bh}\wedge e^{\ch} \bigr)\ ,
\eeq
where the term with $f^i_{\ah\bh}$ satisfies (\ref{3.1}) all by itself (as does $R$) due to the properties of the
coset $G/H$, and the term proportional to $\dot X_{\ah}$ solves (\ref{3.1}) after the proper
choice of $N^{\ah}_{\bh\ch}$ to be specified below.

\medskip

\noindent
{\bf Instantons on conical KT manifolds.} Consider the sine-cone $M^{2n{+}2}=C_\L(\Mcal)$ over a Sasaki-Einstein manifold $\Mcal$ of dimension $2n{+}1$. The geometry of
$\Mcal$ and $M^{2n+2}$ has been discussed in Section 2. In this case, we have $\{\ah\} =\{a, 2n{+}1\}$ with $a=1,\ldots,2n, G=\,$SU$(n{+}1)$, $H=\,$SU$(n)$ and
\beq\label{3.18}
su(n{+}1) \= su(n)\oplus\mfrak\ .
\eeq
The sine-cone $M^{2n+2}$ is conformally equivalent to the cylinder $\Cyl(\Mcal)$ with the local basis 1-forms $e^{\ah}$ and $e^{2n+2}$. The group SO($2n{+}2$)
acts on the tangent spaces of both $\Cyl(\Mcal)$ and $C_\L (\Mcal)$. We have
\beq\label{3.19}
so(2n{+}2)\=su(n{+}1)\oplus u(1)\oplus \Pcal\ ,
\eeq
so the space of antisymmetric $(2n{+}2)\times (2n{+}2)$ matrices can be split into three mutually orthogonal subspaces, which defines $\Pcal$.
The $su(n)$ subspace contains the first two terms in (\ref{3.17}), which are thus $\Sigma$-anti-self-dual.
Using the explicit form of the projector from $so(2n{+}2)$ to $su(n{+}1)$~\cite{16}, one can show that
the subspace $\mfrak$ in (\ref{3.18}) is spanned by the 2-forms (regarded as antisymmetric matrices)
\beq\label{3.20}
e^{2n+1}\wedge e^{2n+2} - \sfrac{1}{2n}\ome_{ab}\,e^a\wedge e^b\and e^a\wedge e^{2n+2}-J^a_b\,e^b\wedge e^{2n+1}\ ,
\eeq
which  satisfy the instanton equations (\ref{3.1}). Here $\ome_{ab}$ and $J^a_b$ are as defined in Section~2. Hence, if we choose $N^{\ah}_{\bh\ch}$
in such a way that the last term in (\ref{3.17}) becomes a linear combination of the 2-forms (\ref{3.20}), then $\Fcal$ from (\ref{3.17}) will also solve the instanton equations (\ref{3.1}).

From (\ref{2.18}), (\ref{2.30}), (\ref{3.17}) and (\ref{3.20}) we finally obtain
\beq\label{3.21}
T^a_{b\,2n+1} = -\sfrac{n+1}{n}\, J^a_b= -f^a_{b\,2n+1} \and T_{ab}^{2n+1} = 2P_{ab\,2n+1}=-2\ome_{ab}= -f_{ab}^{2n+1}\ ,
\eeq
\beq\label{3.22}
N^a_{b\,2n+1} = J^a_b=\sfrac{n}{n+1}\,f^a_{b\,2n+1} \and N_{ab}^{2n+1}= \sfrac{1}{n}\,\ome_{ab}=\sfrac{1}{2n}\,f_{ab}^{2n+1}\ .
\eeq
Substituting (\ref{3.21}) and (\ref{3.22}) into (\ref{3.15}) and (\ref{3.16}), we arrive at
\beq\label{3.23}
[I_i, X_a ]\= f^b_{ia}X_b \and [I_i, X_{2n+1} ]\=0\ ,
\eeq
\beq\label{3.24}
[X_a , X_b ] \= f_{ab}^{2n+1}\bigl(X_{2n+1}+\sfrac{1}{2n}\,\dot X_{2n+1}\bigr)\ +\ f_{ab}^jN_j(\tau)\ ,
\eeq
\beq\label{3.25}
[X_{2n+1} , X_a ] \= f_{{2n+1}\,a}^b \bigl(X_b+\sfrac{n}{n+1}\,\dot X_b\bigr)\ .
\eeq

The task now is to find solutions to the above matrix equations (\ref{3.23})--(\ref{3.25}).
The simplest choice is 
\beq
X_a(\tau) \=\psi(\tau)\, I_a \and X_{2n+1}(\tau) \=\c(\tau)\, I_{2n+1}
\eeq
introducing two functions $\psi$ and $\c$ of $\tau$, which
is related with $r$ via (\ref{2.11}). For this choice, the conditions (\ref{3.23}) are fulfilled, 
and we get $N_i=\psi^2 I_i$. 
This reduces (\ref{3.24})--(\ref{3.25}) to the equations
\beq\label{3.26}
\dot\psi \=\sfrac{n+1}{n}\,\psi\,(\c -1)\=-\frac{\pa W}{\pa\psi} \and 
\dot\c \= 2n\,(\psi^2 -\c )\=-\la^2\,\frac{\pa W}{\pa\c}\ ,
\eeq
which agrees with (4.21) and (4.22) of~\cite{9} for the metric cone. Here, $\la=2n/\sqrt{n{+}1}$, 
and we introduced the flow potential
\beq
W(\psi,\c) \= \sfrac{n{+}1}{2n}\bigl(\psi^2+\sfrac12\c^2-\psi^2\c\bigr)
\eeq
for the variables $\psi$ and $\tilde\c=\c/\la$, so that the second equation in (\ref{3.26}) reads
$\dot{\tilde\c}=-\pa W/\pa\tilde\c$.
For an instanton solution, we need $\psi$ and $\c$ to remain bounded for all $\tau\in\R$.
This requires the flow to start and end in a critical point of~$W$.
Modulo the obvious reflection symmetry $\psi\to-\psi$, the critical points of $W$ are
\beq
\textrm{the local minimum} \quad (\psi,\c)=(0,0) \and 
\textrm{the saddle point} \quad (\psi,\c)=(1,1)\ ,
\eeq
and the flow trajectory connecting them is a separatrix for the vector field $\nabla W$. It is given by
\beq
2\,\psi\,(1-\c)\,\diff\c \= \la^2\,(\c-\psi^2)\,\diff\psi\ ,
\eeq
which admits analytic solutions only for 
\beq
n=1:\ \c=\psi \and n\to\infty:\ \c=\psi^2\ .
\eeq
These and the numerical solutions for $n=2,4,8$ have been plotted in Figure~1 of~\cite{9}.
Here, we display the equipotential lines of the flow potential $W$ and for $n{=}2$ the corresponding streamlines.
The unique bounded solution to (\ref{3.26}) with $\tau$ defined by (\ref{2.11}) yields a Yang-Mills instanton after
substituting $X_a=\psi I_a$ and  $X_{2n+1} =\c I_{2n+1}$ into (\ref{3.12}) and (\ref{3.17}).ß
\begin{figure}[ht]
\centerline{
\includegraphics[width=8cm]{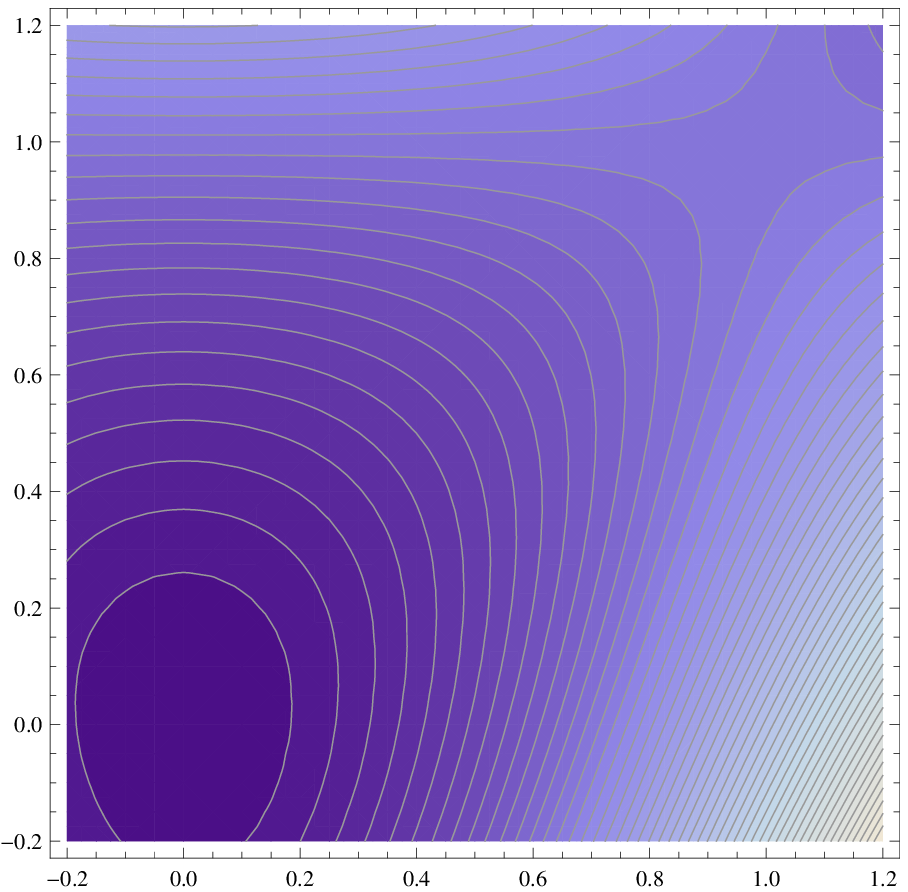}
\hfill
\includegraphics[width=8cm]{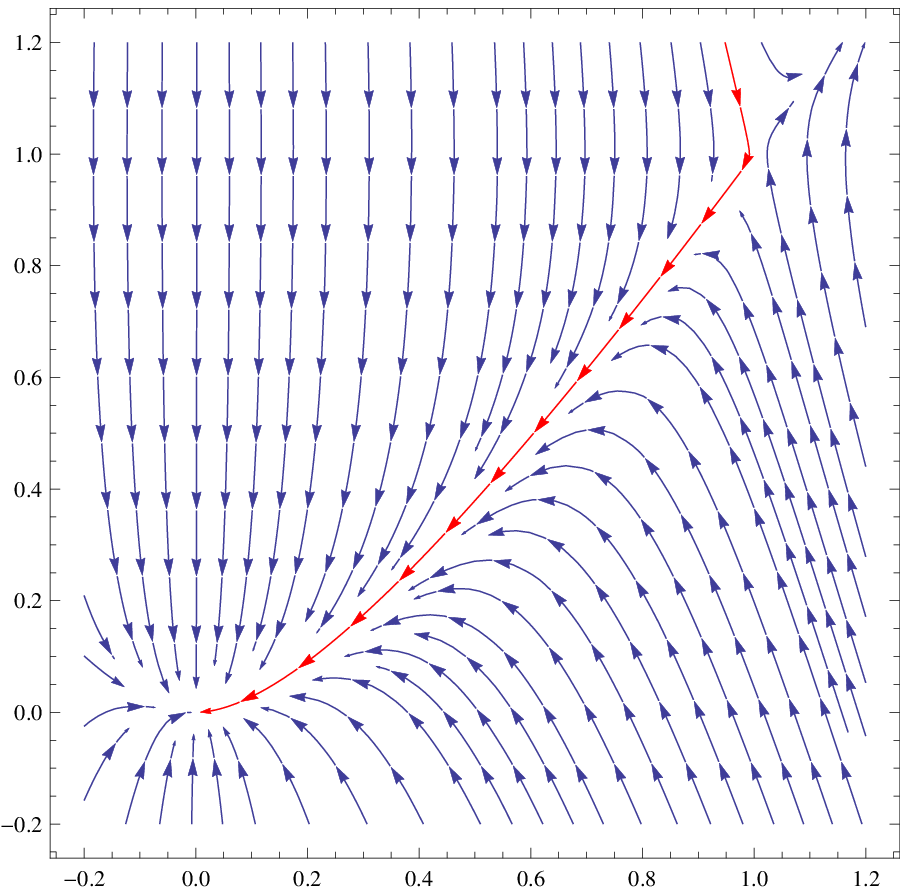}
}
\caption{
equipotential lines for $W(\psi,\c)$ (left) and streamlines for $n{=}2$ (right)}
\label{fig:1}
\end{figure}

\medskip

\noindent
{\bf Instantons on conical HKT manifolds.} In Section~2 we have shown that, for $\Mcal$ being 3-Sasakian of dimension $4n{+}3$, on $M^{4n+4}= C_\L(\Mcal)$ one can introduce a HKT structure with a metric
conformally equivalent to the metric on the cylinder $\Cyl(\Mcal)$. For this reason it suffices to investigate the instanton equation (\ref{3.1}) on $\Cyl(\Mcal)$.
In the 3-Sasakian case, $\{\ah\} =\{a,4n{+}\a\}$ with $a=1,\ldots,4n$ and $\a=1,2,3$, and we have $G=\,$Sp$(n{+}1)$ and $H=\,$Sp$(n)$, i.e.
\beq\label{3.27}
sp(n{+}1)\=sp(n)\oplus\mfrak\ .
\eeq
The group SO$(4n{+}4)$ acts on tangent spaces of both  $\Cyl(\Mcal)$ and $M^{4n+4}$. One gets
\beq\nonumber
so(4n{+}4)\=sp(n{+}1)\oplus sp(1)\oplus\Pcal'\ ,
\eeq
and the $\Sigma$-anti-self-dual 2-forms reside in the $sp(n{+}1)$ subspace of the space of antisymmetric $(4n{+}4)\times (4n{+}4)$ matrices~\cite{16}. Now the first two terms in
(\ref{3.17}) sit in $sp(n)\subset sp(n{+}1)\subset so(4n{+}4)$ and, therefore, satisfy the instanton equation (\ref{3.1}). Employing the explicit form of the projector from $so(4n{+}4)$
to $sp(n{+}1)$~\cite{16}, one can show that the 2-forms
\beq\nonumber
f_2\,\bar\h^\a_{\m\n}e^{4n+\m}\wedge e^{4n+\n}\=f_2\,\bigl(\ve^\a_{\b\g}e^{4n+\b}\wedge e^{4n+\g}-2e^{4n+\a}\wedge e^{4n+4}\bigr)
\eeq
\beq\label{3.28}
\and (f_1f_2)^{1/2} \bigl(e^a\wedge e^{4n+4} + J^{\a a}_{~b}e^b\wedge e^{4n+\a}\bigr)
\eeq
are $\Sigma$-anti-self-dual and form a basis of the subspace $\mfrak$ in (\ref{3.27}).
Here $e^{4n+4}=\diff\tau$, and $J^{\a a}_{~b}$ as well as the functions $f_1, f_2$ have been defined in Section~2. 
From  (\ref{2.27}), (\ref{3.17}) and (\ref{3.28})
it follows that the non-vanishing components are given by
\beq\label{3.29}
T^{4n+\a}_{ab}=-f^{4n+\a}_{ab}\ ,\quad 
T^b_{a\,4n+\b}=-f^b_{a\,4n+\b} \and 
T^{4n+\a}_{4n+\b\, {4n+\g}}=-f^{4n+\a}_{4n+\b\, {4n+\g}}\ ,
\eeq
\beq\label{3.30}
N^a_{b\,4n+\a} = -J^{\a a}_{~b}=-\ome^\a_{ab}=f^a_{b\,4n+\a}\ ,\quad N^{4n+\a}_{4n+\b\, {4n+\g}}=\ve^\a_{\b\g}=\sfrac12\,f^{4n+\a}_{4n+\b\, {4n+\g}}\ .
\eeq
Substituting (\ref{3.29}) and (\ref{3.30}) into (\ref{3.15}) and (\ref{3.16}), we arrive at
\bea\label{3.31}
&& [I_i, X_a ]\= f^b_{ia}X_b \and [I_i, X_{4n+\a} ]\=0\ ,\\[4pt] 
\label{3.32}
&& [X_a , X_b ] \= f_{ab}^{4n+\a}X_{4n+\a}\ +\ f_{ab}^iN_i\ ,\\[4pt]
\label{3.33}
&& [X_a , X_{4n+\b} ] \= f_{a\,{4n+\b}}^b (X_b+\dot X_b)\ , \\[2pt]
\label{3.34} 
&& [X_{4n+\a} , X_{4n+\b} ]\= f_{{4n+\a}\,{4n+\b}}^{4n+\g}
(X_{4n+\g}+\sfrac12\dot X_{4n+\g})\ ,
\eea
again independent of the functions $f_1$ and
$f_2$. If we choose the simplest ansatz 
\beq
X_a(\tau)\=\psi(\tau)\, I_a \and X_{4n+\a}(\tau)\=\c(\tau)\, I_{4n+\a}\ ,
\eeq
then (\ref{3.31}) will be satisfied identically. 
From (\ref{3.32}) we obtain $N_i=\psi^2I_i$, and (\ref{3.33}) and (\ref{3.34}) reduce to 
\beq\label{3.35} 
\dot\psi \=\psi\,(\c -1) \and \dot\c \= 2\,\c\,(\c -1)\qquad\textrm{as well as}\qquad \c =\psi^2\ , \eeq
which is the $n\to\infty$ limit of (\ref{3.26}) and coincides with (4.31)-(4.33) of~\cite{9} for the metric cone.
The equations decouple to
\beq
\dot\psi \= \psi\,(\psi+1)(\psi-1) \and \dot\c \= 2\,\c\,(\c-1)\ ,
\eeq
whose only bounded solution is
\beq\label{3.36} 
\c \= \psi^2 \= \sfrac12\bigl( 1 - \tanh(\tau{-}\tau_0)\bigr) \ .\eeq
Substituting $X_a=\psi I_a$ and
$X_{4n+\a}=\c I_{4n+\a}$ into  (\ref{3.12}) and (\ref{3.17}),
one obtains Yang-Mills instanton configurations on the HKT manifold $M^{4n+4}$ after using
(\ref{2.11}) and the relations  (\ref{2.46}) between the co-frame fields on  $M^{4n+4}$ and the cylinder. More
general instanton solutions may be obtained by considering more general ans\"atze for the matrices $X_{\ah}$.

\bigskip

\section{Conclusions}

\noindent
A Killing spinor on a Riemannian manifold $\Mcal$ is a
spinor field $\epsilon$ obeying the equation $\nabla_{\ah}\epsilon =
\im\lambda\gamma_{\ah}\epsilon$, where $\nabla_{\ah}$ is the spinor
covariant derivative, $\gamma_{\ah}$ are Clifford $\gamma$-matrices
and $\lambda$ is a constant. Manifolds with real Killing spinors often
occur in string-theory compactifications. All these manifolds feature 
connections with non-vanishing torsion and admit a non-integrable
$H$-structure, i.e.\ a reduction of the structure group SO($m$) of the
tangent bundle $T\Mcal$ to $H\subset \ $SO($m$). The metric cone
$C(\Mcal )$ over any such manifold $\Mcal$ has a special (reduced)
holonomy group $G\subset\,$SO$(m{+}1)$ and a Killing spinor $\epsilon$ with
$\lambda =0$ (called parallel spinor). These manifolds were classified in~\cite{35}, 
and, besides the round spheres, they are the \\[-20pt]
\begin{itemize}
\addtolength{\itemsep}{-6pt}
\item nearly K\"ahler 6-manifolds, with $H{=}\,$SU(3) and $G{=}\,G_2$
\item nearly parallel 7-manifolds, with $H{=}\,G_2$ and $G{=}\,$Spin(7)
\item Sasaki-Einstein $(2n{+}1)$-manifolds, with $H{=}\,$SU($n$) and $G{=}\,$SU($n{+}1$) 
\item 3-Sasakian $(4n{+}3)$-manifolds, with $H{=}\,$Sp($n$) and $G{=}\,$Sp($n{+}1$)
\end{itemize}

Instantons on metric cones $C(\Mcal)$ over the above manifolds $\Mcal$ were described in~\cite{9, 22, 23, 26}. Instantons on sine-cones over nearly K\"ahler 6-manifolds and nearly parallel 7-manifolds with $G_2$-structure were investigated in~\cite{25}. 
Here, we completed this study by describing Yang-Mills instantons on sine-cones over Sasaki-Einstein and 3-Sasakian manifolds. In~\cite{9, 10} instantons on metric cones were
extended to brane-type solutions of heterotic supergravity. It would be
of interest to perform a similar lift of instantons on sine-cones. 

\bigskip

\noindent
{\bf Acknowledgements}

\medskip

\noindent
This work was supported in part by the Deutsche
Forschungsgemeinschaft under the grant LE 838/13 and by  the
Heisenberg-Landau program.

\newpage


\begin{thebibliography}{99}

\bibitem{1}
  C.P.~Boyer and K.~Galicki,
  ``3-Sasakian manifolds,''\\
  Surveys Diff.\ Geom.\  {\bf 7} (1999) 123
  [hep-th/9810250].

\bibitem{2}
 C.P.~Boyer and K.~Galicki,
  ``Sasakian geometry, holonomy, and supersymmetry,'' \\ in:
  Handbook of Pseudo-Riemannian Geometry and Supersymmetry, \\
  European Mathematical Society Publishing House 2010, pp.39-84
  [math/0703231 [math.DG]]. 
  
\bibitem{3}
 C.P.~Boyer and K.~Galicki, Sasakian geometry, Oxford University Press, Oxford, 2008.
 
\bibitem{4}
  J.P.~Gauntlett, D.~Martelli, J.~Sparks and D.~Waldram,
  ``Sasaki-Einstein metrics on $S^2 \times S^3$,''\\
  Adv.\ Theor.\ Math.\ Phys.\  {\bf 8} (2004) 711
  [hep-th/0403002];\\
  J.P.~Gauntlett, D.~Martelli, J.F.~Sparks and D.~Waldram,
  ``A new infinite class of Sasaki-Einstein manifolds,''
  Adv.\ Theor.\ Math.\ Phys.\  {\bf 8} (2006) 987
  [hep-th/0403038];\\
   D.~Martelli and J.~Sparks,
  ``Toric Sasaki-Einstein metrics on  $S^2 \times S^3$,''\\
  Phys.\ Lett.\ B {\bf 621} (2005) 208
  [hep-th/0505027].
  
\bibitem{5}
  M.~Cvetic, H.~L\"u, D.N.~Page and C.N.~Pope,
  ``New Einstein-Sasaki spaces in five and higher dimensions,''
  Phys.\ Rev.\ Lett.\  {\bf 95} (2005) 071101
  [hep-th/0504225];\\
  M.~Cvetic, H.~L\"u, D.N.~Page and C.N.~Pope,
  ``New Einstein-Sasaki and Einstein spaces from Kerr-de Sitter,''
  JHEP {\bf 07} (2009) 082
  [hep-th/0505223].
  
\bibitem{6}
  J.~Sparks,
  ``Sasaki-Einstein Manifolds,''\\
  Surveys Diff.\ Geom.\  {\bf 16} (2011) 265
  [arXiv:1004.2461 [math.DG]].
  
\bibitem{7}
  B.S.~Acharya, J.M.~Figueroa-O'Farrill, C.M.~Hull and B.J.~Spence,
  ``Branes at conical singularities and holography,''
  Adv.\ Theor.\ Math.\ Phys.\  {\bf 2} (1999) 1249
  [hep-th/9808014].
  
\bibitem{8}
  P.~Koerber, D.~L\"ust and D.~Tsimpis,
  ``Type IIA AdS$_4$ compactifications on cosets, interpolations and domain walls,''
  JHEP {\bf 07} (2008) 017
  [arXiv:0804.0614 [hep-th]].
  
\bibitem{9}
  D.~Harland and C.~N\"olle,
  ``Instantons and Killing spinors,''\\
  JHEP {\bf 03} (2012) 082
  [arXiv:1109.3552 [hep-th]].
  
\bibitem{10}
  K.-P.~Gemmer, A.S.~Haupt, O.~Lechtenfeld, C.~N\"olle and A.D.~Popov,\\
  ``Heterotic string plus five-brane systems with asymptotic AdS$_3$,''\\
  Adv.\ Theor.\ Math.\ Phys.\  {\bf 17} (2013) 771
  [arXiv:1202.5046 [hep-th]];\\
  C.~N\"olle,
  ``Instantons, five-branes and fractional strings,''
  arXiv:1207.7268 [hep-th].
  
\bibitem{11}
  P.S.~Howe and G.~Papadopoulos,
   ``Twistor spaces for HKT manifolds,''\\
  Phys.\ Lett.\ B {\bf 379} (1996) 80
  [hep-th/9602108].
  
\bibitem{12}
  G.W.~Gibbons, G.~Papadopoulos and K.S.~Stelle,
  ``HKT and OKT geometries on soliton black hole moduli spaces,''
  Nucl.\ Phys.\ B {\bf 508} (1997) 623
  [hep-th/9706207];\\
  A.~Opfermann and G.~Papadopoulos,
  ``Homogeneous HKT and QKT manifolds,''\\
  math-ph/9807026;\\
  J.~Gutowski and G.~Papadopoulos,
  ``The moduli spaces of world volume brane solitons,''\\
  Phys.\ Lett.\ B {\bf 432} (1998) 97
  [hep-th/9802186].
  
\bibitem{13a}
 G.~Grantcharov and Y.S.~Poon,
  ``Geometry of hyper-K\"ahler connections with torsion,''\\
  Commun.\ Math.\ Phys.\  {\bf 213} (2000) 19
  [math/9908015 [math.DG]].

\bibitem{13b}  
  G.~Grantcharov, G.~Papadopoulos and Y.S.~Poon,
  ``Reduction of HKT structures,''\\
  J.\ Math.\ Phys.\  {\bf 43} (2002) 3766
  [math/0201159 [math.DG]].
  
\bibitem{14} 
M.L.~Barberis, ``A survey on hyper-K\"ahler with tortion geometry,''\\
Revista de la Union Math.\ Argentina, {\bf 49} (2008) 121;\\
M.~Fernandes, A.~Fino, L.~Ugarte and R.~Villacampa,\\
``HKT structures from almost contact manifolds,''\\
XIX Int. Fall Workshop on Geometry and Physics, AIP Conf. Proc. {\bf 1360} (2011) 27-38.
  
\bibitem{15}
  E.~Corrigan, C.~Devchand, D.B.~Fairlie and J.~Nuyts,
  ``First order equations for gauge fields in spaces of dimension
    greater than four,''
  Nucl.\ Phys.\ B {\bf 214} (1983) 452;\\
  R.S.~Ward,
  ``Completely solvable gauge field equations in dimension
    greater than four,''\\
  Nucl.\ Phys.\ B {\bf 236} (1984) 381.
  
\bibitem{16}
 T.A.~Ivanova and A.D.~Popov,
  ``(Anti)self-dual gauge fields in dimension $d{\ge}4$,''\\
  Theor.\ Math.\ Phys.\ {\bf 94} (1993) 225.
  
\bibitem{17}
  L.~Baulieu, H.~Kanno and I.M.~Singer,
  ``Special quantum field theories in eight and other dimensions,''
  Commun.\ Math.\ Phys.\ {\bf 194} (1998) 149
  [hep-th/9704167];\\
  M.~Blau and G.~Thompson,
  ``Euclidean SYM theories by time reduction and special holonomy manifolds,''
  Phys.\ Lett.\ B {\bf 415} (1997) 242
  [hep-th/9706225];\\
  B.S.~Acharya, J.M.~Figueroa-O'Farrill, B.J.~Spence and M.~O'Loughlin,
  ``Euclidean D-branes and higher-dimensional gauge theory,''
  Nucl.\ Phys.\  B {\bf 514} (1998) 583
  [hep-th/9707118].

\bibitem{18} 
  S.K.~Donaldson,
  ``Infinite determinants, stable bundles and curvature,''\\
  Duke Math.\ J.\ {\bf 54} (1987) 231;\\
  K.K.~Uhlenbeck and S.-T.~Yau,
  ``On the existence of Hermitian-Yang-Mills connections on stable bundles
    over compact K\"ahler manifolds,''
  Commun.\ Pure Appl.\ Math.\ {\bf 39} (1986) 257.

\bibitem{19}
  S.K.~Donaldson and R.P.~Thomas,
  ``Gauge theory in higher dimensions,''\\
  in:  The Geometric Universe,
  Oxford University Press, Oxford, 1998;\\
 S.K.~Donaldson and E.~Segal,
  ``Gauge theory in higher dimensions II'',\\ 
  Surveys Diff.\ Geom.\ {\bf 16} (2011) 1
  [arXiv:0902.3239 [math.DG]].
  
\bibitem{20}
  G.~Tian,
  ``Gauge theory and calibrated geometry,''\\
  Ann.\ Math.\ {\bf 151} (2000) 193
  [math/0010015 [math.DG]].
 
\bibitem{21}
  D.B.~Fairlie and J.~Nuyts,
  ``Spherically symmetric solutions of gauge theories in eight dimensions,''
  J.\ Phys.\ A {\bf 17} (1984) 2867;\\
 S.~Fubini and H.~Nicolai,
  ``The octonionic instanton,''
  Phys.\ Lett.\ B {\bf 155} (1985) 369;\\
  T.A.~Ivanova and A.D.~Popov,
  ``Self-dual Yang-Mills fields in $d{=}7, 8$, octonions and Ward equations,''
  Lett.\ Math.\ Phys.\  {\bf 24} (1992) 85.
  
\bibitem{22}  
  T.A.~Ivanova, O.~Lechtenfeld, A.D.~Popov and T.~Rahn,
  ``Instantons and Yang-Mills flows on coset spaces,''
  Lett.\ Math.\ Phys.\  {\bf 89} (2009) 231
  [arXiv:0904.0654 [hep-th]];\\
  D.~Harland, T.A.~Ivanova, O.~Lechtenfeld and A.D.~Popov,\\
  ``Yang-Mills flows on nearly K\"ahler manifolds and $G_2$-instantons,''\\
  Commun.\ Math.\ Phys.\ {\bf 300} (2010) 185
  [arXiv:0909.2730 [hep-th]].

\bibitem{23}
  I.~Bauer, T.A.~Ivanova, O.~Lechtenfeld and F.~Lubbe,
  ``Yang-Mills instantons and dyons on homogeneous $G_2$-manifolds,''
  JHEP {\bf 10} (2010) 044
  [arXiv:1006.2388 [hep-th]];\\
 A.S.~Haupt, T.A.~Ivanova, O.~Lechtenfeld and A.D.~Popov,\\
  ``Chern-Simons flows on Aloff-Wallach spaces and Spin(7)-instantons,''\\
  Phys.\ Rev.\ D {\bf 83} (2011) 105028,
  [arXiv:1104.5231 [hep-th]].
  
\bibitem{24}
  F.P.~Correia,
  ``Hermitian Yang-Mills instantons on Calabi-Yau cones,''\\
  JHEP {\bf 12} (2009) 004
  [arXiv:0910.1096 [hep-th]];\\
  F.P.~Correia,
  ``Hermitian Yang-Mills instantons on resolutions of Calabi-Yau cones,''\\
  JHEP {\bf 02} (2011) 054
  [arXiv:1009.0526 [hep-th]].
  
\bibitem{25} 
  K.-P.~Gemmer, O.~Lechtenfeld, C.~N\"olle and A.D.~Popov,
  ``Yang-Mills instantons on cones and sine-cones over nearly K\"ahler
  manifolds,''
  JHEP {\bf 09} (2011) 103
  [arXiv:1108.3951 [hep-th]].

\bibitem{26}   
  T.A.~Ivanova and A.D.~Popov,
  ``Instantons on special holonomy manifolds,''\\
  Phys.\ Rev.\ D {\bf 85} (2012) 105012
  [arXiv:1203.2657 [hep-th]].
 
\bibitem{27}    
  T.~Walpuski,
 ``$G_2$-instantons on generalised Kummer constructions,''\\
  Geom.\ Topol.\  {\bf 17} (2013) 2345
  [arXiv:1109.6609 [math.DG]];\\
  A.~Clarke,
  ``Instantons on the exceptional holonomy manifolds of Bryant and Salamon,''\\
  J.\ Geom.\ Phys.\  {\bf 82} (2014) 84
  [arXiv:1308.6358 [math.DG]].
  
\bibitem{28}      
J.A.~Wolf, Spaces of constant scalar curvature, McGraw-Hill, New York, 1967;\\
J.A.~Wolf and A.~Gray, ``Homogeneous spaces defined by Lie group automorphisms I,II,''\\
J.\ Diff.\ Geom.\ {\bf 2} (1968) 77, 115.

\bibitem{29}     
A.~Gray, ``Nearly K\"ahler manifolds,'' J.\ Diff.\ Geom.\ {\bf 4} (1970) 283.

\bibitem{30}     
P.~Gauduchon, ``Hermitian connections and Dirac operators,''\\ 
Bollettino U.M.I.\ {\bf 11B} (1997) 257;\\
J.M.~Bismut, ``A local index theorem for non-K\"ahler manifolds,'' 
Math.\ Ann.\ {\bf 284} (1989) 681.

\bibitem{31}
A.~Fino, M.~Parton and S.~Salamon, ``Families of strong KT structures in six dimensions,''\\
Comment.\ Math.\ Helv.\ {\bf 79} (2004) 317;\\
M.~Fernandez, A.~Fino, L.~Ugarte and R.~Villacampa, 
``Strong K\"ahler with torsion structures from almost contact manifolds,'' 
Pacific J.\  Math.\ {\bf 249} (2011) 49.

\bibitem{32}
J.~Gutowski, S.~Ivanov and G.~Papadopoulos,
``Deformations of generalized calibrations and compact non-K\"ahler manifolds with vanishing first Chern class,''\\
Asian J.\ Math.\ {\bf 7} (2003) 39 [math/0205012 [math.DG]];\\
D.~Grantcharov, G.~Grantcharov and Y.S.~Poon,
``Calabi-Yau connections with torsion on toric bundles,'' 
J.\ Diff.\ Geom.\ {\bf 78} (2008) 13 [math/0306207 [math.DG]].
  
\bibitem{34}
  M.~Gra\~na,
  ``Flux compactifications in string theory: A comprehensive review,''\\
  Phys.\ Rept.\ {\bf 423} (2006) 91
  [hep-th/0509003];\\
  M.R.~Douglas and S.~Kachru,
  ``Flux compactification,''\\
  Rev.\ Mod.\ Phys.\ {\bf 79} (2007) 733
  [hep-th/0610102];\\
  R.~Blumenhagen, B.~Kors, D.~L\"ust and S.~Stieberger,
  ``Four-dimensional string compactifications with D-branes, orientifolds 
    and fluxes,''
  Phys.\ Rept.\ {\bf 445} (2007) 1
  [hep-th/0610327].

\bibitem{35}
C.~B\"ar, 
``Real Killing spinors and holonomy,''
Commun.\ Math.\ Phys.\ {\bf 154} (1993) 509.

\end{thebibliography}
\end{document}